\documentclass[twocolumn,aps,showpacs,preprintnumbers,amsmath,amssymb, superscriptaddress]{revtex4-1}
\pdfoutput=1
\usepackage{bm}
\usepackage{graphicx}
\usepackage{color}
\usepackage{xspace}
\definecolor{dblue}{rgb}{0,0,0.6}
\definecolor{dred}{rgb}{0.9,0,0}
\definecolor{dgreen}{rgb}{0,0.4,0}

\begin{document}

\title{Raman scattering investigation of large positive magnetoresistance material WTe$_2$}

\author{W. -D. Kong}\thanks{These two authors contributed equally to this work.}
\affiliation{Beijing National Laboratory for Condensed Matter Physics, and Institute of Physics, Chinese Academy of Sciences, Beijing 100190, China}
\author{S. -F. Wu}\thanks{These two authors contributed equally to this work.}
\affiliation{Beijing National Laboratory for Condensed Matter Physics, and Institute of Physics, Chinese Academy of Sciences, Beijing 100190, China}
\author{P. Richard}\email{p.richard@iphy.ac.cn}
\affiliation{Beijing National Laboratory for Condensed Matter Physics, and Institute of Physics, Chinese Academy of Sciences, Beijing 100190, China}
\affiliation{Collaborative Innovation Center of Quantum Matter, Beijing, China}
\author{C. -S. Lian}
\affiliation{Beijing National Laboratory for Condensed Matter Physics, and Institute of Physics, Chinese Academy of Sciences, Beijing 100190, China}
\author{J. -T. Wang}
\affiliation{Beijing National Laboratory for Condensed Matter Physics, and Institute of Physics, Chinese Academy of Sciences, Beijing 100190, China}
\author{C. -L. Yang}
\affiliation{Beijing National Laboratory for Condensed Matter Physics, and Institute of Physics, Chinese Academy of Sciences, Beijing 100190, China}
\author{Y. -G. Shi}
\affiliation{Beijing National Laboratory for Condensed Matter Physics, and Institute of Physics, Chinese Academy of Sciences, Beijing 100190, China}
\author{H. Ding}\email{dingh@iphy.ac.cn}
\affiliation{Beijing National Laboratory for Condensed Matter Physics, and Institute of Physics, Chinese Academy of Sciences, Beijing 100190, China}
\affiliation{Collaborative Innovation Center of Quantum Matter, Beijing, China}

\date{\today}

\begin{abstract}
We have performed polarized Raman scattering measurements on WTe$_2$, for which an extremely large positive magnetoresistance has been reported recently. We observe 5 A$_1$ phonon modes and 2 A$_2$ phonon modes out of 33 Raman active modes, with frequencies in good accordance with first-principles calculations. The angular dependence of the intensity of the peaks observed is consistent with the Raman tensors of the $C_{2v}$ point group symmetry attributed to WTe$_2$. Although the phonon spectra suggest neither strong electron-phonon nor spin-phonon coupling, the intensity of the A$_1$ phonon mode at 160.6 cm$^{-1}$ shows an unconventional decrease with temperature decreasing, for which the origin remains unclear.
\end{abstract}



\maketitle

Giant magnetoresistance is at the core of several important applications, notably for the storage of information. The recent discovery of extremely large positive magnetoresistance (XMR) in layered WTe$_2$ \cite{WTe2Cava} triggered sudden interest for this material. In particular, the non-saturating XMR in WTe$_2$ has been attributed to perfectly balanced electron-hole populations \cite{WTe2Cava,WTe2ARPES}, similar as in pure bismuth and graphite \cite{BismuthAndGraphitePRL2005,GraphitePRL90}. Interestingly, this effect is strongly affected by external pressure \cite{WTe2highPressure}, and pressure-induced superconductivity has even been reported \cite{Arxiv_Pan1501,Arxiv_Kang1502}, which questions the importance of the interactions between the electronic structure and the lattice in WTe$_2$, and offers additional possibilities for development of devices. Unfortunately, literature still lacks of report on the  dynamical properties of the lattice in this system.  


In this letter, we use Raman scattering spectroscopy to characterize the phonons of WTe$_2$ single-crystals. We observe 7 out of 33 Raman active  modes, with frequencies in good accordance with our first-principles calculations. The angular dependence of the Raman intensity of these modes is consistent with their symmetry assignments in terms of the $C_{2v}$ point group symmetry of WTe$_2$. In contrast to our expectation, none of the phonons observed shows evidence for an electron-phonon coupling. However, the intensity of a A$_1$ phonon peak at 160.6 cm$^{-1}$ exhibits an unusual decrease upon cooling, whose origin remains unclear.

The WTe$_2$ single crystals used in our Raman scattering measurements were grown by solid-state reactions. The resistivity of the samples was measured with a Quantum Design physical properties measurement system (PPMS). The crystals were cleaved in air to obtain flat surfaces and then transferred into a low-temperature cryostat ST500 (Janis) for the Raman measurements between 5 and 300 K with a working vacuum better than $8\times 10^{-7}$ mbar. Raman scattering measurements were performed using a 514.5 nm excitation laser in a back-scattering micro-Raman configuration, with a triple-grating spectrometer (Horiba Jobin Yvon T64000) equipped with a nitrogen-cooled CCD camera. In this manuscript, we define $x$ and $y$ as the directions along the $a$ axis (W-W chains) and $b$ axis, respectively. $x'$ and $y'$ are oriented at 45$^{\circ}$ from the $x$ and $y$. The $z$ direction corresponds to the $c$ axis perpendicular to the W-Te planes.

The  WTe$_2$ crystal structure is characterized by the space group \textsl{Pmn2$_1$} ($C_{2v}^{7}$, No.31) \cite{Mar_JACS114}. A simple group symmetry analysis \cite{Comarou_Bilbao} indicates that the phonon modes at the Brillouin zone (BZ) center decompose into
[11A$_1$+6A$_2$+5B$_1$+11B$_2$]+[11A$_1$+5B$_1$+11B$_2$]+ [A$_1$+B$_1$+B$_2$], where the first, second and third terms represent the Raman-active modes, the infrared(IR)-active modes and the acoustic modes, respectively. To get estimates on the phonon frequencies, we performed first-principles calculations of the phonon modes at the BZ center in the framework of the density functional perturbation theory (DFPT) \cite{Baroni_RMP73}, using the experimental lattice parameters $a=3.496$ \AA\xspace, $b=6.282$ \AA\xspace and $c=14.0730$ \AA. The Wyckoff positions of all the atoms are 2a. For all calculations, we used the Vienna \emph{ab initio} simulation package (VASP) \cite{KressePRB54} with the generalized gradient approximation (GGA) of Perdew-Burke-Ernzerhof for the exchange-correlation functions \cite{Perdew_PRB46}. The projector augmented wave (PAW) \cite{Blochl_PRB50} method was employed to describe the electron-ion interactions. A plane wave cutoff energy of 520 eV was used with a uniform $9\times 9\times 9$ Monkhorst-Pack $k$-point mesh for integrations over the BZ. The frequencies of the phonon modes were derived from the dynamical matrix generated by the DFPT method. The calculated frequencies and the experimental phonon modes are given in Table \ref{EXP_CAL_comparsion}.

\begin{figure}[!t]
\begin{center}
\includegraphics[width=3.4in]{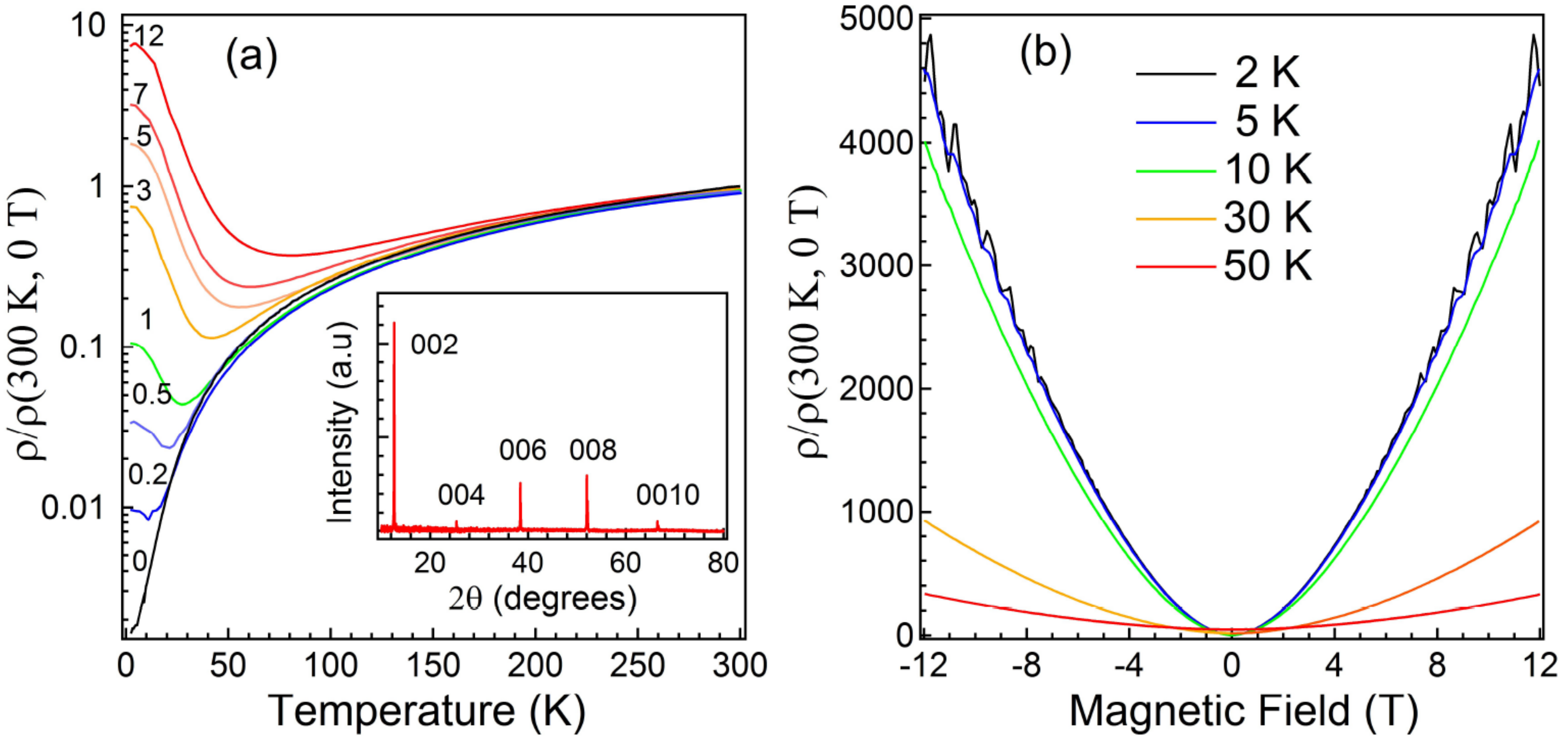}
\end{center}
\caption{\label{transport}(Color online). Temperature and field dependence of the XMR in WTe$_2$, with the current parallel to \textsl{a} and \textsl{H} parallel to \textsl{c}. (a) Normalized resistivity (at 300 K and 0 T) as a function of temperature, for different magnitudes of the magnetic field. The labels above the curves correspond to the applied field in teslas. Inset: XRD data, showing high single-crystal quality. (b) Magnetoresistance of WTe$_2$ at different temperatures.}
\end{figure}

In Fig. \ref{transport}(a), we show the  temperature-dependent resistivity of our WTe$_2$ samples under various applied magnetic fields ($H$ up to 12 T). In agreement with previous reports \cite{WTe2Cava,WTe2highPressure}, the magnetoresistance at low temperature is extremely large, corresponding to an increase factor reaching 4538 at 2 K in a field of 12 T. In Fig. \ref{transport}(b), we show the field dependence of the XMR at various temperatures. Shubnikov-de Haas quantum oscillations are clearly observed at 2 K, and we checked that the frequency spectra is compatible with the ones reported previously \cite{WTe2highPressure}, thus suggesting the good quality of our samples. Our samples quality is also suggested by the sharp peaks observed in the x-ray diffraction (XRD) data shown the inset in Fig. \ref{transport}(a).

\begin{figure}[!t]
\begin{center}
\includegraphics[width=3.4in]{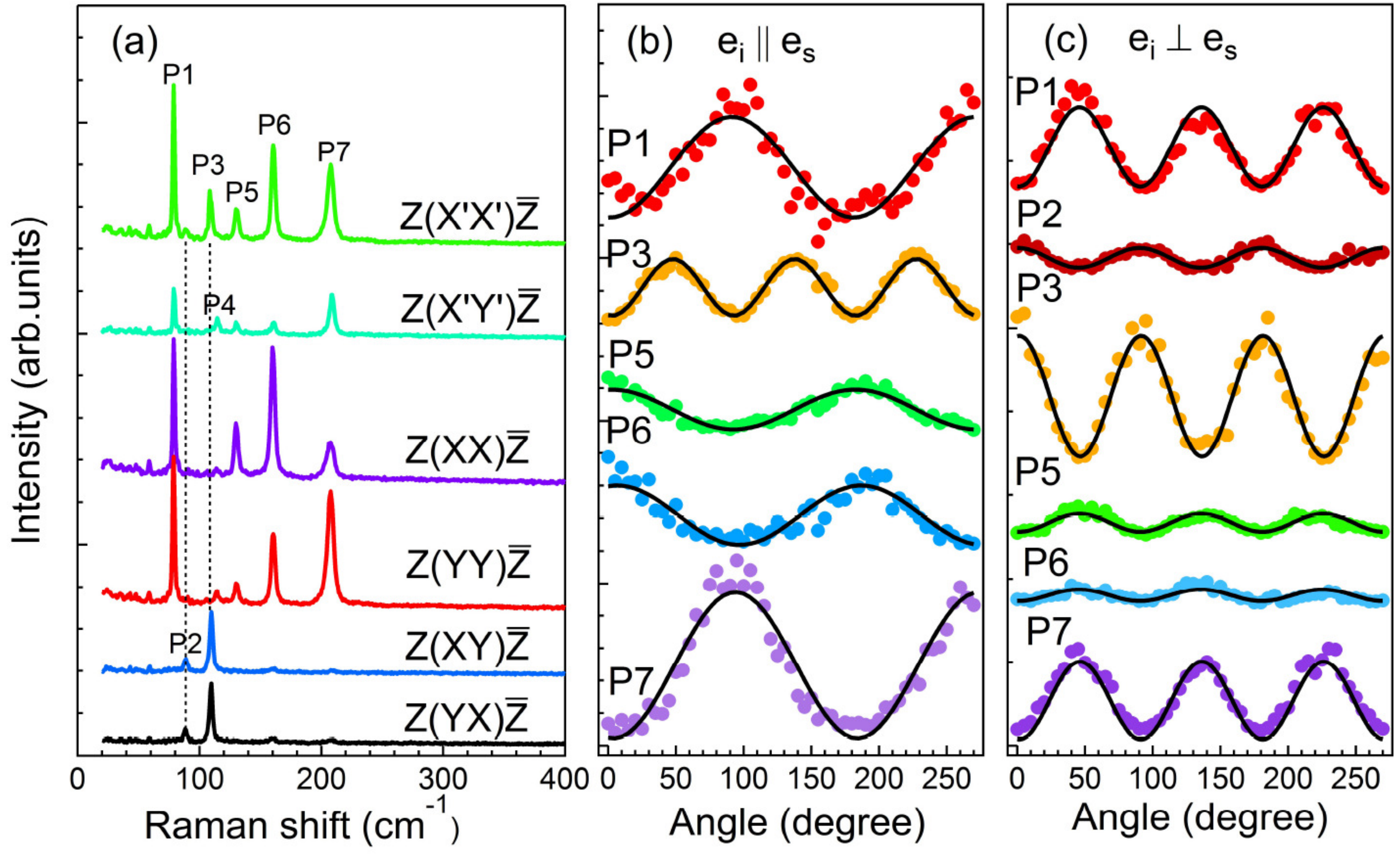}
\end{center}
\caption{\label{phonons_fig}(Color online). (a) Polarization dependence of the $ab$ plane measurements on WTe$_2$ at 294 K. The Raman modes are identified by labels running from P1 to P7. The vertical dashed lines are guides to the eye for the P2 and P3 peaks. (b) Angle dependence (defined with respect to $x$) of the phonons (labeled according to (a)) recorded at 294 K with $e_{i}|| e_{s}$ in the $ab$ plane. The intensity as been shifted upward for sake of clarity. The black lines are guides to the eye. (c) Same as (b) but for $e_{i}\perp e_{s}$. }
\end{figure}

In Fig. \ref{phonons_fig}(a), we show the Raman spectra of WTe$_2$ recorded at room temperature under various polarization configurations. The symmetry of the modes observed are determined by the Raman tensors corresponding to the $C_{2v}$ symmetry group, which are expressed as:

\begin{displaymath}
\textrm{A$_1$}=
\left(\begin{array}{ccc}
a & 0 &0\\
0 & b &0\\
0 & 0 &c
\end{array}\right)
, \textrm{A$_{2}$ =}
\left(\begin{array}{ccc}
0 & d &0\\
d & 0 &0\\
0 & 0 &0\\
\end{array}\right),
\end{displaymath}

\begin{displaymath}
\textrm{B$_1$}=
\left(\begin{array}{ccc}
0 & 0 &e\\
0 & 0 &0\\
e & 0 &0
\end{array}\right)
, \textrm{B$_{2}$}=
\left(\begin{array}{ccc}
0 & 0 &0\\
0 & 0 &f\\
0 & f &0\\
\end{array}\right).
\end{displaymath}
\noindent 
While both A$_{1}$ and A$_{2}$ modes can be detected for in-plane configurations of the incident ($e_{i}$) and scattered ($e_{s}$) polarization vectors, B$_{1}$ and B$_{2}$ modes can be observed only if either $e_{i}$ or $e_{s}$ is aligned along the $c$ axis. Therefore, the latter two symmetries are forbidden in our experimental configuration. For perfectly aligned crystals, pure A$_{1}$ symmetry is obtained in the $z(xx)\bar{z}$ and $z(yy)\bar{z}$ configurations. In these two configurations, we detect 5 sharp peaks at 78.9 cm$^{-1}$ (P1), 114.6 cm$^{-1}$ (P4), 129.9 cm$^{-1}$ (P5), 160.6 cm$^{-1}$ (P6) and 207.7 cm$^{-1}$ (P7). The relative intensity of the peaks differs in the $z(xx)\bar{z}$ and $z(yy)\bar{z}$ spectra, as expect from the two-fold symmetry of the $ab$ plane of this material, as also expressed by the Raman tensor A$_1$. Pure A$_{2}$ symmetry is obtained in the equivalent $z(xy)\bar{z}$ and $z(yx)\bar{z}$ configurations. The corresponding spectra exhibit two sharp peaks at 88.4 cm$^{-1}$ (P2) and 109.9 cm$^{-1}$ (P3). When the $e_{i}$ and $e_{s}$ polarization vectors are not along the $a$ or $b$ axes, the symmetry of the spectra is no longer pure and both A$_{1}$ and A$_{2}$ phonons are detected. For example, all the peaks labeled P1 to P7 are observed in the $z(x'x')\bar{z}$ and $z(x'y')\bar{z}$ channels. As indicated by the comparison given in Table \ref{EXP_CAL_comparsion}, the experimental frequencies of the phonons P1 to P7 is quite well reproduced by our calculations, and the corresponding symmetry assignments match perfectly.   

\begin{table}
\caption{\label{EXP_CAL_comparsion}Comparison of the calculated and experimental of Raman active phonon modes (in cm$^{-1}$) at 294 K. IR indicates infrared activity.}
\begin{ruledtabular}
\begin{tabular}{ccccc}
 Symmetry &	Activity &	  Experiment &	Calculation	& \\
\hline
A$_{1}$&    Raman+IR&             &       8.9&  \\
A$_{2}$& 	Raman& 	              &	     24.4&	\\
B$_{2}$& 	Raman+IR& 	          &      28.7&	\\
A$_{1}$& 	Raman+IR& 	     78.9 (P1)&	     75.7&	\\
B$_{2}$& 	Raman+IR& 		      &      85.9& 	\\
B$_{1}$& 	Raman+IR& 	          &      87.3&	\\
A$_{2}$&    Raman& 	          88.4 (P2)&      89.1& 	\\
A$_{2}$& 	Raman& 	         109.9 (P3)&     113.2& \\
B$_{1}$& 	Raman+IR& 	          &     113.9&\\
A$_{1}$& 	Raman+IR&        114.6 (P4)&     115.2&	\\
A$_{2}$& 	Raman& 	              &     117.3&	\\
B$_{2}$& 	Raman+IR& 		      &     119.3& \\
B$_{1}$& 	Raman+IR& 	          &     119.4&	\\
B$_{2}$& 	Raman+IR& 	          &     127.7& \\
B$_{2}$& 	Raman+IR& 	          &     131.9&\\
A$_{1}$& 	Raman+IR& 	     129.9 (P5)&     132.0&	\\
A$_{1}$& 	Raman+IR& 		      &     135.0&\\
B$_{2}$& 	Raman+IR& 		      &     135.1&\\
A$_{1}$& 	Raman+IR& 		      &     136.3&\\
A$_{2}$& 	Raman& 		          &     157.6&	\\
B$_{1}$& 	Raman+IR& 		      &     157.9&	\\
B$_{2}$& 	Raman+IR& 		      &     161.6&	\\
A$_{2}$& 	Raman& 		          &     163.6&	\\
B$_{1}$& 	Raman+IR& 		      &     164.7&	\\
A$_{1}$& 	Raman+IR&        160.6 (P6)&     165.7&	\\
B$_{2}$& 	Raman+IR& 		      &     178.6&	\\
A$_{1}$& 	Raman+IR& 		      &     179.5&	\\
A$_{1}$& 	Raman+IR& 	     207.7 (P7)&     211.3&	\\
B$_{2}$& 	Raman+IR& 		      &     211.8&	\\
A$_{1}$& 	Raman+IR& 		      &     215.8&	\\
B$_{2}$& 	Raman+IR& 		      &     217.6&	\\
A$_{1}$& 	Raman+IR& 		      &     240.3&	\\
B$_{2}$& 	Raman+IR& 		      &     240.7&	\\

\end{tabular}
\end{ruledtabular}
\begin{raggedright}
\end{raggedright}
\end{table}

In Figs. \ref{phonons_fig}(b) and (c), we show the in-plane angle dependence of the phonon peak intensities in the parallel and perpendicular configurations of the $e_{i}$ and $e_{s}$ polarization vectors, respectively. As expected from the $C_{2v}$ symmetry, a four-fold symmetry is observed in the perpendicular configuration for all peaks (see Fig. \ref{phonons_fig}(c)). However, the phase of oscillations associated to the intensities of the P2 and P3 peaks differs by $\pi$ from that of the other peaks. In the parallel configuration (Fig. \ref{phonons_fig}(b)), all the A$_1$ peak exhibit a two-fold symmetry, as predicted by the A$_1$ tensor. Since their intensity never vanishes, we conclude that the diagonal terms of the A$_1$ tensor are not equal. We also note that the oscillations of the intensity of the P1 and P7 peaks are in anti-phase with those of the P5 and P6 peaks. In contrast to the intensity of the A$_1$ peaks, the intensity of the P3 peak shows a four-fold symmetry. This behavior is easily understood in terms of the A$_2$ Raman tensor, with the peak intensity vanishing when $e_{i}$ and $e_{s}$ are parallel to either $a$ or $b$. We note that this perfect agreement between our experiment and the predictions from the crystal symmetry, especially in the case of the peak P3, contrasts with the anomalous angular data reported in a recent Raman study \cite{Arxiv_Jiang1501}. We attribute this discrepancy to a better sample quality in our experiments and by the distortion of the spectra induced by the use of a filter in Ref. \cite{Arxiv_Jiang1501}.   

To investigate the possible role of the electron-phonon or phonon-phonon interactions on the XMR at low temperature, we cooled the samples down to 5 K. In Fig. \ref{T_dependent}(a), we display the temperature dependence of the A$_{1}$ peaks at 114.6 cm$^{-1}$ (P5), 129.9 cm$^{-1}$ (P6) and 160.6 cm$^{-1}$ (P7), which have the strongest intensities in the $z(xx)\bar{z}$ polarization channel. As expected, the peaks become a little sharper with decreasing temperature. The symmetric Lorentzian lineshapes at all temperatures (see Fig. \ref{T_dependent}(i) for an example) suggest that there is neither strong electron-phonon coupling nor spin-phonon coupling in this system, at least for the phonon modes probed by our experiments. We show in Figs. \ref{T_dependent}(b-d) and \ref{T_dependent}(e-g) a quantitative analysis of the peak positions and linewidths of the P5, P6 and P7 peaks, which have been fit simultaneously with three Lorentzian functions convoluted by a Gaussian function representing the system resolution. In each case, the peak position $\omega_{ph}(T)$ and the linewidth $\Gamma_{ph}(T)$ follows simple expressions corresponding to the anharmonic phonon decay into acoustic phonons with the same frequencies and opposite momenta \cite{Klemens_PhysRev148,Menendez_PRB29,wu2014BaTiFeAsO,wu2014BiS2}:

\begin{figure}[!t]
\begin{center}
\includegraphics[width=3.4in]{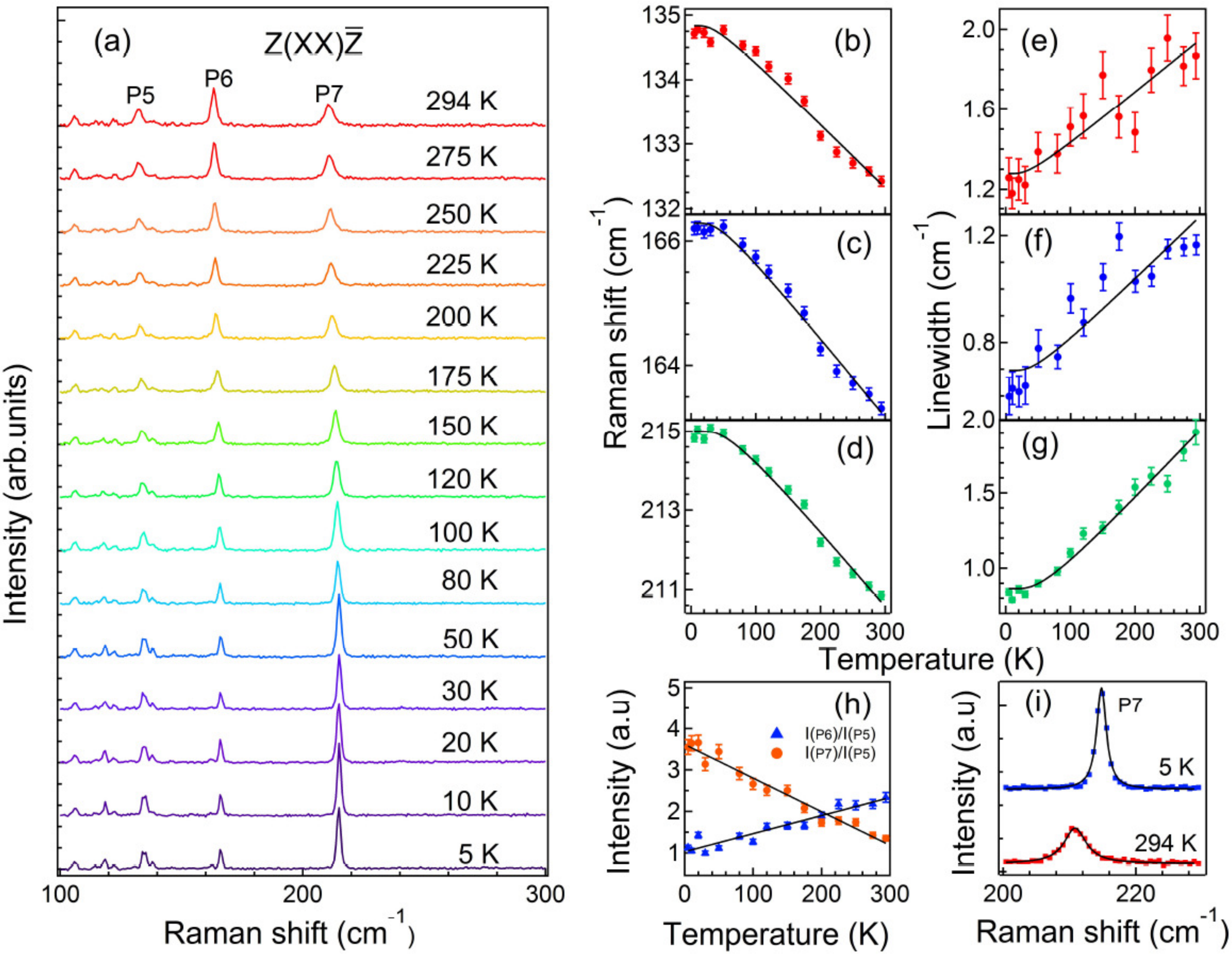}
\end{center}
\caption{\label{T_dependent}(Color online). (a) Temperature dependence of the Raman spectra for the $z(xx)\bar{z}$ configuration.  (b)-(d)  Temperature dependence of peak positions of P5, P6 and P7 defined in Fig. \ref{phonons_fig}(b). (e)-(g) Temperature dependence of the corresponding linewidths (full-width-at-half-maximum) of P5, P6 and P7, respectively. (h) Temperature dependence of the intensity ratios of the P6 and P5 peaks [I(P6)/I(P5)], and of the P7 and P5 peaks [I(P7)/I(P5)]. The black lines are linear fits to the data. (i) Zoom on the P7 peak at 5 K and 294 K.}
\end{figure}

\begin{equation}
\label{eq_omega}
\omega_{ph}(T)=\omega_{0}-C\left( 1+\frac{2}{e^{\frac{\hbar\omega_0 }{ 2k_{B}T}} -1} \right )
\end{equation}
\begin{equation}
\label{eq_gamma}
\Gamma_{ph}(T)=\Gamma_{0}+\Gamma\left( 1+\frac{2}{e^{\frac{\hbar\omega_0 }{ 2k_{B}T}} -1} \right ),
\end{equation}

\noindent where $C$ and $\Gamma$ are positive constants, $\omega_0$ is the bare phonon frequency, and $\Gamma_0$ is a residual, temperature-independent linewidth. The fitting parameters are given in TABLE \ref{Fit_para}.
 \begin{table}[!t]
\caption{\label{Fit_para}Fitting parameters (in cm$^{-1}$) for the peak positions and linewidths}
\begin{ruledtabular}
\begin{tabular}{cccccc}
 peak &	$\omega$ &	$C$ &	$\Gamma$ & $\Gamma_0$&\\
\hline
P$5$&    135.32&    0.482&  1.15&   0.128&\\
P$6$&    167.05&    0.767&  0.553&  0.142&\\
P$7$&    216.51&    1.506&  0.506&  0.358&\\
\end{tabular}
\end{ruledtabular}
\begin{raggedright}
\end{raggedright}
\end{table}

While the intensities of all the other peaks increase on cooling, which is particularly true for the P7 peak, the intensity of the P6 peak exhibits an unusual decrease, as displayed in Fig. \ref{T_dependent}(a). To emphasize this point we show in Fig. \ref{T_dependent} (h) the intensity ratios of the P6 and P5 peaks [I(P6)/I(P5)] and of the P7 and P5 peaks [I(P7)/I(P5)]. The temperature variations are linear in both cases. Whether the unconventional behavior of the P6 peak can be related to the strong temperature dependence of the resistivity upon cooling, is unclear.

In summary, we have performed polarized Raman scattering measurements on WTe$_2$, for which a large positive magnetoresistance has been reported recently. We observe seven (5A$_1$+2A$_2$) out of 33 Raman active modes, with frequencies in good accordance with first-principles calculations. The intensity of these peak as a function of the in-plane angular polarization is consistent with the $C_{2v}$ point group symmetry attributed to WTe$_2$. The phonon spectra suggest neither strong electron-phonon nor spin-phonon coupling. and the temperature dependence of A$_1$ phonon peak positions and linewidths have been analyzed fit to the standard model of the anharmonic decay of optical phonons. However, we observed an unexpected decrease in the intensity of the A$_1$ phonon at 160.6 cm$^{-1}$ upon cooling.

This work was supported by grants from MOST (2010CB923000, 2011CBA001000, 2011CBA00102, 2012CB821403 and 2013CB921703) and NSFC (11004232, 11034011/A0402, 11234014, 11274362 and 11474330) from China.

\bibliography{biblio_long}

\end{document}